\documentclass[prl,twocolumn,showpacs]{revtex4}
\usepackage[dvipdfm]{graphicx}
\usepackage{amsmath,amssymb,stmaryrd}
\newcommand{\be}{\begin{equation}}
\newcommand{\ee}{\end{equation}}

\begin{document}

\title{Comb entanglement in quantum spin chains}
\author{J.P. Keating}
\email{j.p.keating@bristol.ac.uk}
\author{F. Mezzadri}
\email{f.mezzadri@bristol.ac.uk}
\author{M. Novaes}
\email{marcel.novaes@bristol.ac.uk} \affiliation{School of
Mathematics, University of Bristol, Bristol BS8 1TW, UK}

\begin{abstract}
Bipartite entanglement in the ground state of a chain of $N$ quantum
spins can be quantified either by computing pairwise concurrence or
by dividing the chain into two complementary subsystems. In the
latter case the smaller subsystem is usually a single spin or a
block of adjacent spins and the entanglement differentiates between
critical and non-critical regimes. Here we extend this approach by
considering a more general setting: our smaller subsystem $S_A$
consists of a {\it comb} of $L$ spins, spaced $p$ sites apart. Our
results are thus not restricted to a simple `area law', but contain
non-local information, parameterized by the spacing $p$.  For the XX
model we calculate the von-Neumann entropy analytically when
$N\rightarrow \infty$ and investigate its dependence on $L$ and $p$.
We find that an external magnetic field induces an unexpected length
scale for entanglement in this case.
\end{abstract}


\maketitle

\section{Introduction}

Quantum phase transitions at zero temperature correspond to a
fundamental restructuring of a system's ground state. In quantum
spin chains these transitions occur as an external parameter (e.g.~a
magnetic field) is varied \cite{book}, and are manifested as a
marked change in the decay of quantum correlations: algebraic at the
critical point and exponential away from it. The amount of
entanglement present in the ground state is expected to depend
significantly on whether the system is critical or not, since at a
critical point all the constituent parts of the system must be
non-locally correlated and thus entangled
\cite{pra66tjo2002,n416ao2002,prl90gv2003,qic4jil2004,pra71vp2005,pra71mfy2005}.

Unfortunately, it is not yet clear how to measure entanglement in
general \cite{pra70rs2004}. At present, we only understand fully
how to quantify bipartite entanglement \cite{pra53chb1996}.  It is
natural, therefore, to try to extract as much information as
possible about entanglement in this context. Ways of doing this
have recently been the focus of considerable attention.  One
possibility is the following. From a chain of $N$
spin-$\frac{1}{2}$ particles select two, compute their reduced
density matrix, and then obtain the associated concurrence
\cite{prl80wt1998}. This is a function of the separation between
the selected spins.  Despite the fact that the concurrence is only
a short-range measure (it vanishes if the spins are farther apart
than next-nearest neighbors), this approach has been applied to
detect phase transitions in a variety of situations
\cite{pra66tjo2002,n416ao2002,pra71vp2005}. A second possibility
is to measure the entanglement between a single spin and the rest
of the chain.  This has also been related to the presence of a
critical point \cite{pra66tjo2002}.

The problem with these methods, which involve a small number of
spins, is that they do not take into account the fact that
entanglement is shared between many parties, i.e.~they provide
little information regarding the non-local nature of entanglement.
This deficiency is shared by another much-studied bipartite
division of the spin chain, namely that between a {\it block} of
$L$ {\it adjacent} spins and the remaining $N-L$
\cite{prl90gv2003,qic4jil2004}. In the limit $N\rightarrow\infty$,
the entanglement entropy has in this case been computed
analytically using the theory of Toeplitz determinants (for the XX
model) \cite{jsp116bqj2004,fisher}, conformal field theory
\cite{prl92vek2004}, and from averages over ensembles of random
matrices \cite{cmp252jpk2004}. For one-dimensional chains it has
been shown that as $L\rightarrow \infty$ the entropy tends to a
constant value away from critical points, and that it diverges
like $\ln L$ at phase transitions. For critical $d$-dimensional
spin-lattices this `block' entanglement has been proven to grow
like $L^{d-1}\ln L$ under certain conditions \cite{prl96mmw2006},
while for a gapped system one expects an area scaling law due to
the finite correlation length \cite{cramer}. Such a direct
relation between entanglement and area is known to hold in
harmonic lattices \cite{pra73mc2006}. All these results indicate
that, at least for large blocks, the entanglement comes mostly
from the boundary.

\begin{figure}[b]
\includegraphics[scale=0.39,angle=-90,bb=140 390 400 405]{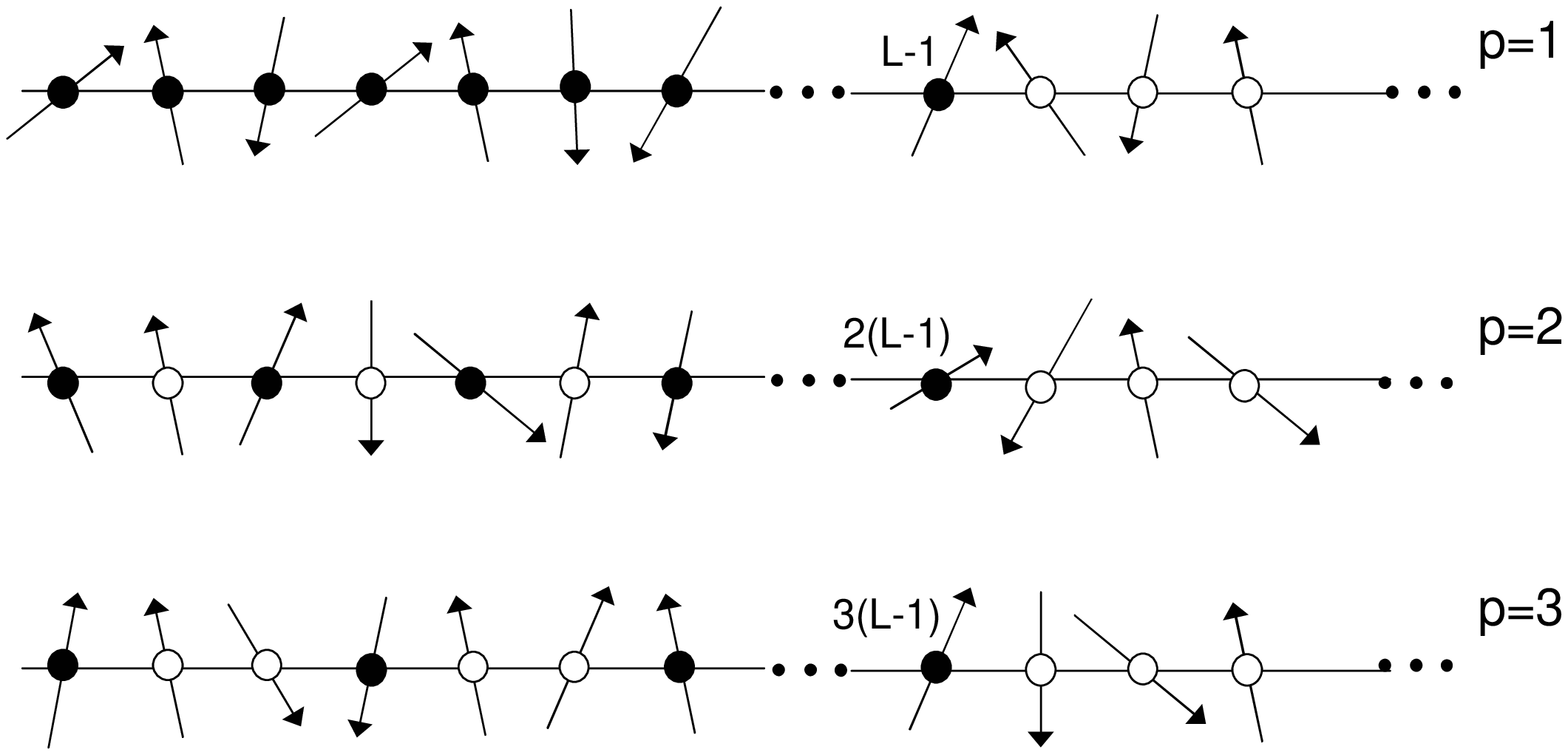}
\caption{The comb division illustrated for three different values of
the spacing $p$. In all cases one subsystem ($S_A$, denoted by black
circles) contains $L$ spins and the other ($S_B$, denoted by empty
circles) contains the rest of the chain. We have written explicitly
the label of the last of the $L$ spins (the label of the first spin
is $0$). The first case, $p=1$, corresponds to the well known
`block' division.}
\end{figure}

Our purpose here is to introduce a new geometry in which to study
bipartite entanglement in quantum spin chains. We divide the chain
into two subsystems, $S_A$ and $S_B$, as follows.  $S_A$ consists of
$L$ equally spaced spins, such that the spacing between the spins in
this subsystem corresponds to $p$ sites on the chain. $S_B$ then
contains the remaining $N-L$ spins. (Obviously this only makes sense
if $N>(L-1)p$.)  $S_A$ can be visualized as a {\it comb} with $L$
teeth. This geometry enables us to study non-local entanglement
effects by varying the spacing $p$.

Three possible divisions are illustrated in Fig.1. For $p=1$ we
recover the simple `block' arrangement, which, as was noted above,
has already been the subject of extensive investigation. In this
case the subsystems $S_A$ and $S_B$ are only `in contact' near
their common border, and as a result, for large values of $L$, the
dependence of the entanglement on $L$ is minimal. For $p>1$ the
entanglement between $S_A$ and $S_B$ is shared between many
different sites and it then grows linearly with $L$, to leading
order as $L\rightarrow \infty$.  The $\log L$ term that dominates
when $p=1$ appears as a secondary correction when $p>1$.  As an
example, we investigate how the comb entanglement in the XX model
depends on the spacing $p$ when $L$ is fixed and show that this
reveals the emergence of a new length-scale determined by the
external magnetic field.

\section{Block entropy}

The XY spin chain in an external uniform magnetic field $h$ has as
its Hamiltonian \be\label{hamil}
H=\sum_{j=0}^{N-1}\left(\frac{1+\gamma}{2}\,
\sigma_j^x\sigma_{j+1}^x+
\frac{1-\gamma}{2}\,\sigma_j^y\sigma_{j+1}^y\right)-h\sum_{j=0}^{N}\sigma_j^z,\ee
where $(\sigma_j^x,\sigma_j^y,\sigma_j^z)$ are the usual Pauli
matrices. When $\gamma=0$ this system is called the XX model and
when $\gamma=1$ it is the Ising model. It is an integrable model
\cite{pra2eb1970} which displays both critical and non-critical
regimes.  In the $\gamma-h$ phase diagram the lines $h=\pm 1$ and
the segment $\gamma=0,\lvert h \rvert <1$ are critical, while all
other regions are non-critical. For simplicity, we will restrict
our analysis mainly to the XX model and will come back to the
general case towards the end.

Since the XX ground state is just a (non-entangled) ferromagnet
for $\lvert h \rvert >1$, we will only consider its critical
regime. A more convenient parameter in that case is the angle $k$
defined by \be h=\cos k, \quad k\in[0,\pi].\ee  We denote by
$|\Psi\rangle$ the ground state of this system and introduce the
Jordan-Wigner (JW) transform at each site of the lattice, \be
m_{2l+1}=\left(\prod_{j=0}^{l-1}\sigma_j^z\right)\sigma_l^x, \quad
m_{2l}=\left(\prod_{j=0}^{l-1}\sigma_j^z\right)\sigma_l^y. \ee
Given any product of an odd number of JW operators, we can see
from the symmetry of the Hamiltonian that its expectation value
with respect to $|\Psi\rangle$ vanishes. Wick's theorem and the
relation $\langle\Psi|m_jm_k|\Psi\rangle=\delta_{jk}+i(C_N)_{jk}$,
where $C_N$ is called the correlation matrix, allow for the
calculation of the expectation value of a product of any number of
JW operators.

The matrix $C_N$ factorizes into a direct product, \be
C_N=T[g]\otimes\left(
                                 \begin{array}{cc}
                                   0 & 1 \\
                                   -1 & 0 \\
                                 \end{array}
                               \right),\ee
where $T[g]$ is the matrix \be (T[g])_{jk}=\tilde{g}_{j-k}, \quad
\tilde{g}_n=\frac{1}{2\pi}\int_0^{2\pi}g(\theta)e^{-in\theta}d\theta.\ee
The function $g(\theta)$ is called the symbol of the Toeplitz
matrix $T[g]$.  For the XX chain it is given by \be\label{g}
g(\theta)=\begin{cases} 1 &\text{if }-k\leq\theta<k,\\-1
&\text{otherwise}.\end{cases}\ee

Following the calculation presented in \cite{jsp116bqj2004}, the
entropy of subsystem $S_A$ is obtained as a contour integral in
the complex plane: \be\label{integ}
E=\lim_{\epsilon\rightarrow0^+}\lim_{\delta\rightarrow0^+}\frac{1}{2\pi
i}\oint_{c(\epsilon,\delta)}e(1+\epsilon,\lambda)\frac{d \ln
D_A(\lambda)}{d\lambda}d\lambda,\ee where $D_A(\lambda)={\rm
det}(\lambda I-T_A[g])$ involves the matrix $T_A[g]$, which is
obtained from the original matrix $T[g]$ by removing the rows and
columns that correspond to sites in $S_B$, and \be
e(x,y)=-\frac{x+y}{2}\log_2\left(\frac{x+y}{2}\right)-
\frac{x-y}{2}\log_2\left(\frac{x-y}{2}\right).\ee The contour of
integration $c(\epsilon,\delta)$ approaches the interval $[-1,1]$
as $\epsilon$ and $\delta$ tend to zero without enclosing the
branch points of $e(1+\epsilon,\lambda)$.

When $S_A$ corresponds to $L$ consecutive spins, i.e.~in the {\it
block} case, $T_A$ is simply a block inside $T$ and thus is also a
Toeplitz matrix. This allowed Jin and Korepin \cite{jsp116bqj2004}
to obtain the corresponding entropy by using a proved instance of
the Fisher-Hartwig conjecture relating to the asymptotics of
Toeplitz determinants \cite{fisher}. This states that in the limit
of large blocks we have \be \ln D_A(\lambda)= c_0L+\beta^2\ln
L+O(1), \ee where \be c_0(\lambda)=\frac{1}{2\pi}\int_0^{2\pi}\ln
\bigl(\lambda-g(\theta)\bigr)d\theta.\ee The coefficient
$\beta(\lambda)$ may be calculated by writing the symbol as \be
\lambda-g(\theta)=\phi(\lambda)t_{\beta}(\theta-k)t_{-\beta}(\theta+k),\ee
with $t_{\beta}(\theta)=e^{-i\beta(\pi-\theta)}$ and \be
\phi(\lambda)=(\lambda+1)\left(\frac{\lambda+1}{\lambda-1}\right)^{-k/\pi},\ee
and hence it is given by \be\beta(\lambda)=-\frac{1}{2\pi
i}\ln\left(\frac{\lambda+1}{\lambda-1}\right).\ee

As a consequence of the Fisher-Hartwig conjecture the entanglement
as $L\rightarrow \infty$ is \be
E(L)=\mathcal{E}_1L+\mathcal{E}_2\ln L+O(1).\ee Because of the
simplicity of the function $g(\theta)$ all the relevant quantities
can be evaluated. In particular, we have \be
c_0(\lambda)=\frac{1}{\pi}\bigl[k\ln(\lambda-1)+(\pi-k)\ln(\lambda+1)\bigr]
,\ee and by substituting this into (\ref{integ}) we see that the
leading order term $\mathcal{E}_1$ actually vanishes because \be
e(1,1)=e(1,-1)=0. \ee
 Therefore the dependence upon $L$ is only logarithmic, with a
prefactor $\mathcal{E}_2$ which is obtained from (\ref{integ}). It
turns out that when $p>1$ the leading order contribution does not
vanish.  The next section is devoted to its computation.

\section{Comb entropy}

The choice we propose for the subsystem $S_A$ also leads to a
Toeplitz structure for the matrix $T_A$. We single out those spins
whose label is a multiple of an integer $p$, and thus we have \be
(T_A)_{jk}=\tilde{g}_{pj-pk}=\frac{1}{2\pi}\int_0^{2\pi}g(\theta)e^{-ip(j-k)\theta}d\theta.\ee
This is not yet in the Toeplitz form. We must first find a
function $g_p(\theta)$ such that \be\label{gtoh}
\int_0^{2\pi}g(\alpha)e^{-ipn\alpha}d\alpha=\int_0^{2\pi}g_p(\alpha)e^{-in\alpha}d\alpha,\ee
and this will be the symbol of $T_A$. Multiplying (\ref{gtoh}) by
$e^{in\theta}$ with $0\leq\theta<2\pi$, summing over $n$ and using
the Poisson summation formula we arrive at \be\label{aver}
g_p(\theta)=\frac{1}{p}\sum_{n=0}^{p-1}g\left(\frac{\theta}{p}+\frac{2n\pi}{p}\right),\ee
so the value of $g_p$ at the point $\theta$ is obtained as the
average value of $g$ over the $p$ vertices of a regular polygon.

This average is not hard to calculate. It is easy to see that for
each $k$ the function $g_p(\theta)$ is piecewise constant and
even, with jumps at the critical points $\pm\theta^\ast$ given by
\be [0,\pi)\ni\theta^\ast=\min \{\alpha,2\pi-\alpha\}, \quad
\alpha=pk \; \bmod 2\pi.\ee Its values are \be\label{aga}
g_p(\theta)=\frac{2}{p}-1+\frac{4}{p}\left\llbracket
\frac{pk}{2\pi}\right\rrbracket+\begin{cases}0&\text{if }
-\theta^\ast\leq\theta<\theta^\ast\\2s/p&\text{otherwise
},\end{cases}\ee where the brackets $\llbracket\cdot\rrbracket$
denote the integer part and \be s={\rm sign}\{\alpha-\pi\}.\ee

The entanglement will now depend on the spacing: \be
E(L;p)=\mathcal{E}_1(p)L+\mathcal{E}_2(p)\ln L+O(1)\ee as $L
\rightarrow \infty $.  To calculate it to leading order we only
need the integral \be
c_0=\frac{1}{2\pi}\int_0^{2\pi}\ln\bigl(\lambda-g_p(\theta)\bigr)d\theta,\ee
and this leads to \be\label{general}
\mathcal{E}_1(p)=\pi^{-1}\bigl[\theta^\ast
e(1,g_p(0))+(\pi-\theta^\ast)e(1,g_p(\pi))\bigr],\ee which is our
main result. We calculate the logarithmic correction
$\mathcal{E}_2(p)$ in the next section. Note that the bound $0\leq
\mathcal{E}_1(p)\leq 1$ is explicitly respected.

If we restrict our subsystem to be a single spin, i.e. $L=1$, it
is easy to calculate the entanglement because the correlation
matrix has only one element,
$(2\pi)^{-1}\int_0^{2\pi}g(\theta)d\theta$. We get simply \be
E_1=e\Bigl(1,\tfrac{2k}{\pi}-1\Bigr).\ee On the other hand, from
equation~\eqref{aga}  we obtain that as $p \rightarrow \infty$
\begin{subequations}
\label{gfunctions}
\begin{align}
g_p(0) & \sim \frac{2k}{\pi} - 1 + q_1(p), \\
\intertext{and} g_p(\pi) & \sim \frac{2k}{\pi} - 1 + q_2(p),
\end{align}
\end{subequations}
where both $q_1(p)$ and $q_2(p)$ vanish like $p^{-1}$. If now we
insert~\eqref{gfunctions} into the general expression
(\ref{general}), then we find that the total entanglement in the
limit of large spacing converges, as expected, to a combination of
single-spin contributions, \be\label{limit} \mathcal{E}_1(p) \sim
E_1+ \frac{a}{p}, \ee where $a$ is a constant. It is interesting
to note that the rate of convergence is rather slow, indicating a
long-range dependence of the entanglement on the spacing. This is
in contrast with the behavior of simpler quantities like the
pairwise concurrence, for example, which vanishes if the spins are
more than two sites apart.

\begin{figure}[t]
\includegraphics[scale=0.4]{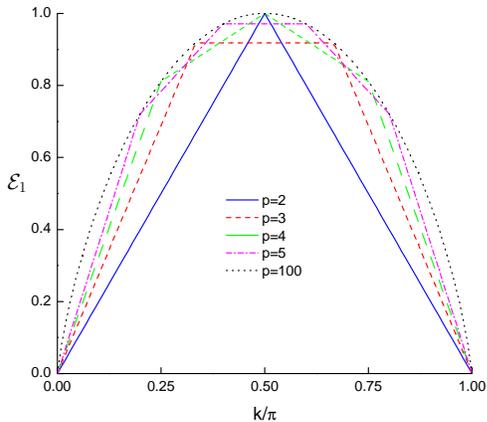}
\caption{(color online) $\mathcal{E}_1$ as a function of $k$ for
various values of the spacing $p$. The curve is always piecewise
linear and weakly converges to $E_1$. Notice that for $k=\pi/2$
the entanglement is maximal whenever $p$ is even.}
\end{figure}

We next consider the case when $p=2$. The values $g_2(0)$ and
$g_2(\pi)$ are given by
\begin{subequations}
\begin{align} g_2(0)& =\frac{1}{2}\bigl (g(0)+g(\pi)\bigr)=0 \\
\intertext{and}
 g_2(\pi)& = \frac{1}{2}\bigl(g(\pi /2)+g(3\pi
/2)\bigr)=g(\pi/2).
\end{align}
\end{subequations}
The latter is either $1$ or $-1$, and thus makes no contribution
to entanglement. The critical angle is just $\theta^\ast=2k$ and
hence we have $\mathcal{E}_1(2)=2Lk/\pi.$ The linear dependence on
$k$ can be seen in Fig.2, where we plot the entanglement for
various values of the spacing $p$. In the absence of any external
magnetic field, i.e. for $k=\pi/2$, the entanglement attains its
maximum possible value whenever $p$ is even. For larger values of
$p$ the function is always piecewise linear, eventually converging
to $E_1$.

It is important to observe that the expression (\ref{general}) for
the entanglement is continuous when we consider $p$ as a real
number, despite the discontinuities that appear in (\ref{aga}).
The function $g_p(0)$ is discontinuous whenever $pk=2n\pi$, but at
those points $\theta^\ast$ vanishes and hence $\mathcal{E}_1(p)$
remains unaffected. On the other hand, $\mathcal{E}_1(p)$ is also
oblivious to the jumps in the function $g_p(\pi)$, which occur at
$pk=(2n+1)\pi$ (due to the variable $s$), because then we have
$\theta^\ast=\pi$. Its derivative, on the other hand, is
discontinuous: at the special points $pk=n\pi$ the entanglement
has a local maximum with a cusp form. Remarkably, for a fixed $k$
its values at these maxima are all the same (i.e.~they do not
depend on $n$) and are equal to the large-$p$ limiting value, \be
\mathcal{E}_1(n\pi/k)=E_1.\ee

In Fig.3 we plot $\mathcal{E}_1(p)$ as a function of the spacing
for different values of the magnetic field $h=\cos k$. Since, of
course, only integer values of $p$ may be realized in the actual
chain, we take $k=\pi/\ell$, where $\ell$ is an integer. We see
that this leads to the appearance of an unexpected length scale
for the entanglement: it attains its maximal value whenever the
spacing is a multiple of $\ell$. The existence of such a length
scale appears to be a fundamental property of quantum spin chains
in magnetic fields.

\begin{figure}[t]
\includegraphics[scale=0.4]{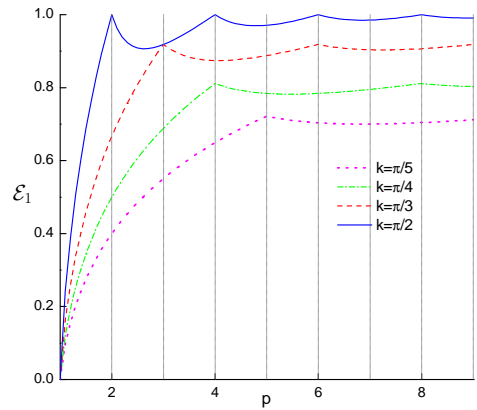}
\caption{(color online) $\mathcal{E}_1$ as a function of the
spacing $p$ for various values of the magnetic field $h=\cos k$.
For $k=\pi/\ell$ there exists a typical length scale for the
spacing: the entanglement is maximal for $p=n\ell$, and its value
at these points is equal to the limit $E_1$.}
\end{figure}

Recently the formalism of Toeplitz determinants has been used to
compute the `block' entanglement of the more general XY model
\cite{jpa38ari2005}, with finite anisotropy parameter $\gamma\neq0$.
The main difference with respect to the case $\gamma=0$ is the form
of the function $g(\theta)$, which is no longer piecewise constant
and becomes complex. Calculating the average (\ref{aver}) then
becomes much less simple, but it is still possible to employ the
present approach to obtain the entanglement for arbitrary values of
the spacing $p$. A special case for which explicit calculations are
possible is the Ising model without magnetic field, obtained from
(\ref{hamil}) by setting $\gamma=1$ and $h=0$. In this case we have
$g(\theta)=e^{i\theta}$ and thus $g_p(\theta)=0$, leading to the
result that $E_{\rm Ising}(L;p)=L$ for any value of $p$. This
reflects the fact that the zero temperature ground state of this
model is maximally entangled \cite{pra66tjo2002}.

\section{Logarithmic correction}

In order to obtain the logarithmic correction for the XX model we
need to decompose the symbol (\ref{gtoh}) in the form \be
\lambda-g_p(\theta)=\phi(\lambda)t_{\beta}(\theta-\theta^\ast)t_{-\beta}(\theta+\theta^\ast),\ee
where \be
\phi(\lambda)=(\lambda-g_p(\pi))\left(\frac{\lambda-g_p(\pi)}{\lambda-g_p(0)}\right)^{-\theta^\ast/\pi}\ee
is independent of $\theta$ and the discontinuities are accounted
for by \be t_\beta(\theta)=e^{-i\beta(\pi-\theta)}, \quad
\theta\in[0,2\pi).\ee The function $\beta(\lambda)$ is given by
\be \beta(\lambda)=\frac{-1}{2\pi
i}\ln\left(\frac{\lambda-g_p(\pi)}{\lambda-g_p(0)}\right),\ee and
the coefficient in the logarithmic correction to the entanglement
is \be \mathcal{E}_2(p)=\frac{1}{\pi i}\oint
e(1,\lambda)\beta(\lambda)\frac{d \beta}{d\lambda}d\lambda.\ee

Since for $p>1$ both $\left \lvert g_p(0)\right \rvert$ and $\lvert
g_p(\pi)\rvert$ are smaller than unity, we can choose as our contour
of integration the unit circle. Using power series expansions we
arrive at \be
\mathcal{E}_2(p)=\frac{g_p(0)-g_p(\pi)}{2\pi^2\ln2}I(g_p(0),g_p(\pi)),\ee
where\be \label{infini}I(a,b)=\sum_{m=0}^\infty \sum_{n=1}^\infty
\sum_{j=1}^n\frac{a^nb^m-a^mb^n}{j(n+m+1)(n+m)}.\ee

Since both $g_p(0)$ and $g_p(\pi)$ are piecewise constant as
functions of $k$, the same is true for $\mathcal{E}_2$. In the
left panel of Fig.4 we see that the number of jumps in this
function grows as $p$ increases. On the other hand, for a given
value of the magnetic field $\mathcal{E}_2$ decays rapidly as $p$
increases, and has discontinuities at $p=n\ell$ when $k=\pi/\ell$
(see the right panel of Fig.4). Notice that the vanishing of the
logarithmic term in the entanglement is consistent with the fact
that $E\rightarrow LE_1$ as $p\to\infty$ (cf.~\eqref{limit}).

\begin{figure}[t]
\includegraphics[scale=0.43]{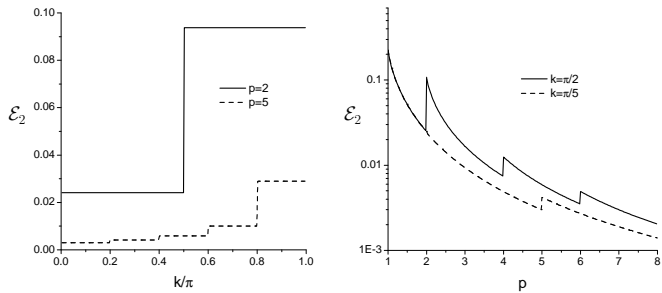}
\caption{Left: $\mathcal{E}_2$ as a function of $k$ for two values
of the spacing. Right: $\mathcal{E}_2$ as a function of $p$ for two
values of the magnetic field.}
\end{figure}

\section{Conclusions}

In summary, we have extended the bipartite approach to entanglement
in one-dimensional critical spin chains by introducing a new
partition of the chain, the {\it comb partition}, which allows us to
go beyond the simple `block' picture and investigate non-local
correlations analytically. The organizing subsystem consists of $L$
spins separated by $p$ sites, and we have found that as
$p\rightarrow\infty$ its entanglement with the rest of the chain
reduces to the sum of the individual contributions of its elements,
although with a slow convergence rate that indicates the existence
of long-range correlations. We have also found that the presence of
a magnetic field induces a typical length scale for entanglement. It
would be interesting the see if this length scale is present in
other statistical properties of critical spin chains. Our results
regarding a generalized version of the Emptiness Formation
Probability, which has recently been computed for the XY model using
Toeplitz determinants \cite{jpa38ff2005}, will appear elsewhere
\cite{sepf}.

We are grateful to Noah Linden for a stimulating discussion. MN
thanks CAPES for financial support.  JPK is supported by an EPSRC
Senior Research Fellowship.

\end{document}